\documentclass[journal]{iopart}

\usepackage[pdftex]{graphicx}

\begin{document}

\note[Practical Way to Measure Large-Scale 2D Surfaces Using Repositioning on CMMs]{
Practical Way to Measure Large-Scale 2D Surfaces Using Repositioning on Coordinate-Measuring Machines}

\author{Sergey~Kosarevsky}

\address{Saint-Petersburg Institute of the Machine-building, Department of Technology, Saint-Petersburg, Russia}
\ead{kosarevsky@mail.ru}

\begin{abstract}
In many situations it is required to perform an inspection of large flat parts
on a coordinate-measuring machine (CMM) when it is impossible
to probe all necessary surfaces from one position of the part. Or
it is necessary to measure a large part which dimensions exceed
the volume of an available CMM. For this purpose one needs to
merge the data measured in two different positions of the workpiece
into one coordinate system. Though most of geodesic software
has out-of-the-box functionality to do that,
a lot of popular CMM software lacks it. In this paper a practical approach is
described to bring a repositioning functionality into
a CMM software. The Calypso metrology software was studied. The proposed 
inspection method can be used both for the measurements of linear 
dimensions and location tolerances as common practice in Calypso.
\end{abstract}

%\begin{keywords}
%Coordinate-measuring machine, Calypso, large parts, repositioning.
%\end{keywords}

\maketitle

% main text
\section{Introduction}

Repositioning on CMMs using reference spheres was first analyzed by 
M.G. Cox \cite{Cox97}. By designing tapped holes into the workpiece where 
reference spheres can be attached, measurements from the two positions can 
be stitched into one coordinate system. A FORTRAN software implementation
for that method was provided by Butler, Forbes and Kenward 
in \cite{Butler98}. But it is too complex for practical everyday usage
with proprietary CMM software and requires strong mathematical and
programming skills from end users.

Here we propose a practical method to reposition flat parts which does not 
require any design changes to the part to be made. We will implement
it using Calypso metrology software. 

\begin{figure}[b]
\begin{center}
   \includegraphics[width=8.5cm]{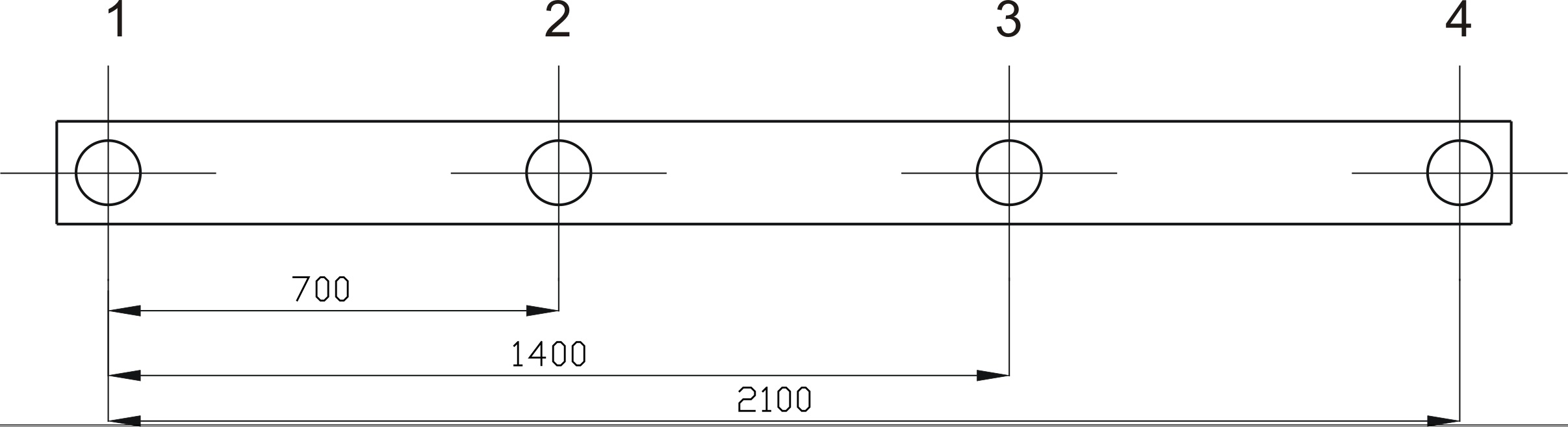} 
   \caption{Sketch of the workpiece}
   \label{Workpiece}
\end{center}
\end{figure}

For clarity let us start 
directly with an application of this method and describe the procedure on a 
concrete example. Imagine a  rectangular plate with four in-line 
holes (fig.~\ref{Workpiece}).

Let us consider it is necessary to measure distances between the center of 
the first hole and the centers of every other hole. Taking into consideration
that measuring range of our particular CMM along the longest axis\footnote{We don't consider diagonal clamping of the part since it 
requires additional fixtures to be designed.} is limited 
to 1500 mm, it is impossible to do the measurement
without moving the part because CMM cannot probe all
the holes in one position. Most of the geodesic software
can do it \cite{Leica97}, \cite{Warren97}, but typical CMM software
does not have out-of-the-box solution for this task. 

We assume that the reader is not only familiar
with the Calypso metrology software, but also has some experience with
parametric coded measurements: Calypso PCM. For a detailed specification of
Calypso please refer to the user guide \cite{CalypsoMan}, and 
for a description of parametric coded measurements see the PCM 
training manual \cite{PCMMan}.

\section{Manual approach}

\begin{figure}
\begin{center}
   \includegraphics[width=6.5cm]{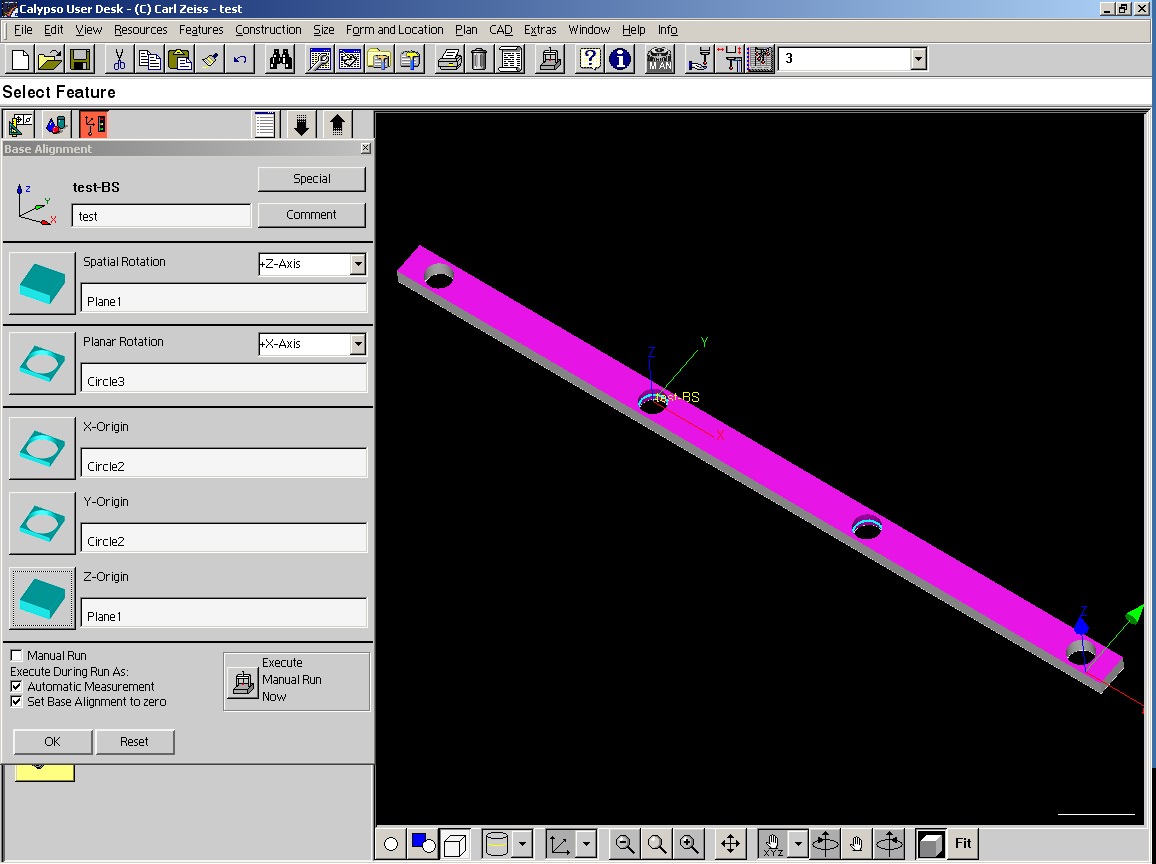}
   \caption{Base alignment properties}
   \label{BaseAlignment}
\end{center}
\end{figure}

We have to bring the coordinates of every hole center into one coordinate 
system in order to solve the abovementioned problem. For this purpose we 
use centers of the holes 2 and 3 as registration points to merge data from 
two positions of the part. Initially, we need to
create a Base alignment for the part in Calypso. Let us use the workpiece's
top plane (plane of the drawing fig.~\ref{Workpiece}) for spatial rotation. We move origin 
of the reference frame to the center of hole 2. Hole 3 is used for the 
planar rotation to define the direction of X axis. So the Base alignment
looks as follows (fig.~\ref{BaseAlignment}): \\

\begin{itemize}
   \item X axis passes through the centers of the holes 2 and 3, from left to right;
   \item Z axis is orthogonal to the top plane of the workpiece, pointing to the viewer;
   \item Y axis is orthogonal to X and Z, pointing down;
   \item XY planar origin is at the center of hole 2;
   \item Z planar origin is at the top plane of the workpiece. \\
\end{itemize}

Let us put the workpiece inside the CMM measuring range so, that
we will be able to measure holes 1, 2 and 3 directly. This is possible
considering the CMM measuring range is limited to 1500 mm.

We can measure the first three holes in CNC mode. We have to 
exclude the fourth holes from CNC run by manually selecting only the features
related to the first three holes (BaseAlignment and Position1 groups)
and specifying ``Current Selection'' before CNC start. After this 
measurement we write down the coordinates of the first hole into a notepad 
--- we will need them later.

Now we are ready to reposition the workpiece and measure the fourth hole. 
First, let us move the workpiece so that the holes 2, 3 and 4 become
accessible by the probing system. Exclude hole 1 from measurement 
(by selection of BaseAlignment and Position2 groups) and restart CNC run 
with manual alignment. After alignment is done the fourth hole is measured 
in CNC mode.

Unfortunately, Calypso purged the measured coordinates of the feature ``Circle1''.
But we can easily restore them\footnote{Since the stored coordinates are in Base Alignment we don't need
to perform any additional transformations.}
from the notepad and place into a theoretical
feature (Circle1').

We can now use the created theoretical feature to create all the 
characteristics we need to determine the required distances. Calypso
evaluates them as if all the measurements had been done in one 
measurement run.

This method successfully measures almost any flat workpiece, in which the 
number of features for realignment is one or two.

\section{PCM approach}

If there are a lot of features to be converted from one alignment
to another the amount of manual work will be incredibly large and
the manual approach described above becomes impractical. Besides,
the manual stage of the method is prone to operator's errors. 
It is a good idea to make the realignment process fully-automatic.

We can use Calypso's parametric coded measurement functionality\footnote{Users without the
Calypso PCM option will be unable to perform actions described in this
section.} to pass the results between two alignments. The general 
workflow remains the same; the only difference is that we  
use a disk file instead of a notepad. We write and read
data to and from this file using a tiny PCM program.

Three following functions from the Calypso PCM programming 
language are necessary to meet our demands for file access: \\

\begin{itemize}
   \item deleteFile( "file name" ) --- delete a file
   \item addToFile(  "file name", "text line" ) --- add a line to a text file; if the file does not exist it will be created
   \item readPCMFile( "file name" ) --- read values of PCM variables from a text file into Calypso\footnote{Those familiar with
                   C/C++ programming languages will find it similiar to include-files.}. \\
\end{itemize}

The operational sequence of our PCM program will be as follows:
delete the file during the first CNC run and write the coordinates
of the first hole into it; read the file during the second CNC run and
apply the stored values to a theoretical feature. First and second CNC
runs refer to first and second position of the workpiece respectively.

The PCM program to be executed after the first CNC run is:

{
\footnotesize
\begin{verbatim}

   deleteFile( "C:\TempFile.param" )

   XStr = format(getActual("Circle1").x)
   YStr = format(getActual("Circle1").y)

   PosXStr = "PositionX = " + XStr
   PosYStr = "PositionY = " + YStr

   addToFile( "C:\TempFile.param", PosXStr )
   addToFile( "C:\TempFile.param", PosYStr )

\end{verbatim}
}

Content of the temporary file after this run is:

{
\footnotesize
\begin{verbatim}

   PositionX = -699.9657d
   PositionY = -0.0138d

\end{verbatim}
}

The PCM program to be executed before the second CNC run is simpler 
and consists of the only line:

\begin{verbatim}

   readPCMFile( "C:\TempFile.param" )

\end{verbatim}

We have to use variables PositionX and PositionY as an X and Y parameters
of the theoretical feature Circle1' respectively. Finally the manual 
part of the work has been eliminated --- all the reading now is done 
automatically.

\section{Registration points}

\begin{figure}
\begin{center}
   \includegraphics[width=8.5cm]{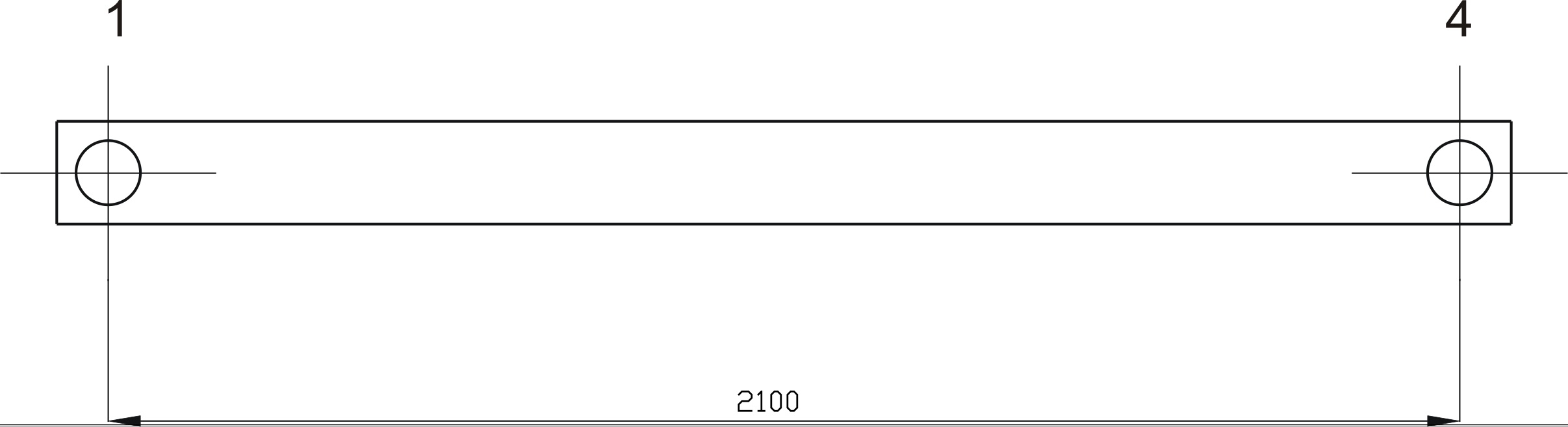}
   \caption{A workpiece without registration points}
   \label{WorkpieceRP}
\end{center}
\end{figure}

In some workpieces there are no features that can be used for 
registration points (same way like holes 2 and 3 were used). An
example for such a workpiece can be obtained by a simple modification
of our original sketch (fig.~\ref{WorkpieceRP}).

\begin{figure}
\begin{center}
   \includegraphics[width=6.5cm]{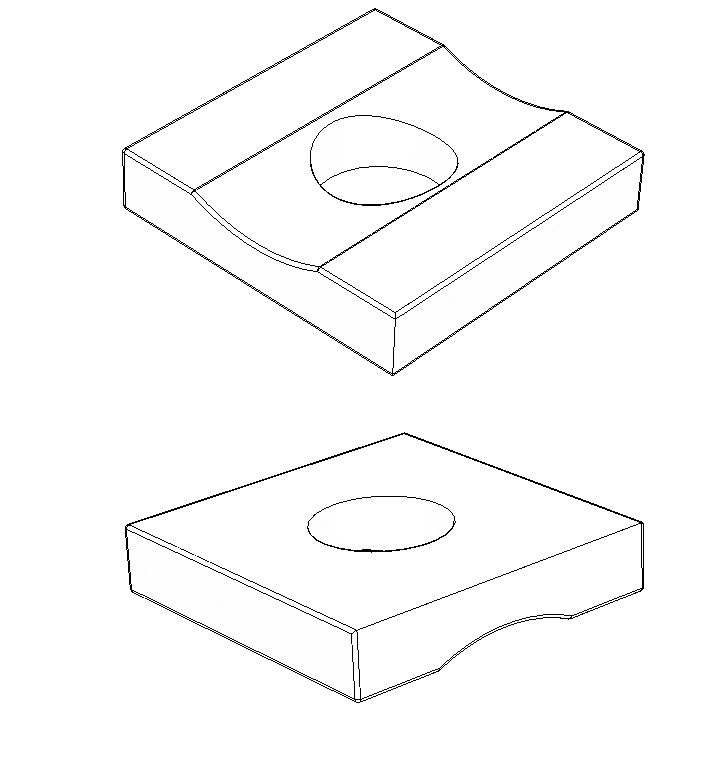}
   \caption{The plate to create a registration point}
   \label{Plate}
\end{center}
\end{figure}

Artificial registration points should be created to overcome that difficulty. 
Cox \cite{Cox97} suggests usage of zirconium spheres for this purpose.
In case of flat parts a simple artifact can be used (fig.~\ref{Plate}).

\begin{figure}
\begin{center}
   \includegraphics[width=8.5cm]{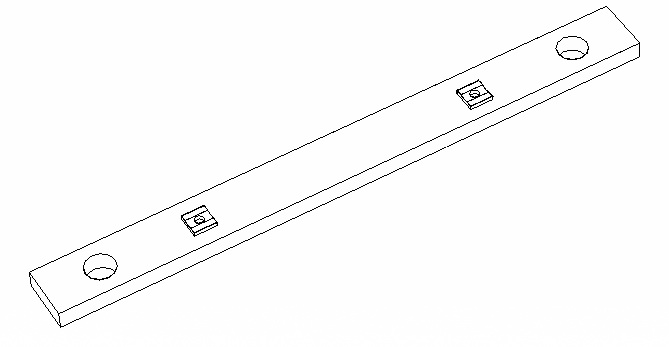}
   \caption{A pair of plates fixed on the workpiece}
   \label{Datum}
\end{center}
\end{figure}

Two of these plates can be attached (i.e. glued or clamped) to the most of
planar or radial surfaces of a workpiece to create an artificial datum which moves
rigidly with the workpiece, i.e. as in fig.~\ref{Datum}. Two-sided scotch tape can 
be efficiently used as a fixture if uncertainty of 0.01 mm is enough.

The distance between the two plates on the workpiece should be kept maximal
(it is limited to the size of the workpiece and the measuring range of CMM)
to prevent creation of a degenerate datum. Both plates
should actually remain intact during both CNC runs. This solution
allows complex parts with freeform surfaces and without any regular geometry 
features to be measured.

\section{Reliability of the method}

Though we do not know how to obtain reliable uncertainty estimates from Calypso, 
it is easy to see that our procedure does not introduce significant additional 
errors. Indeed, we can check it by means of comparison. We measured a small 
block with holes, similar to that at fig.~\ref{Workpiece}, but its length was only 100 mm. 
It was measured with and without repositioning. Deviation of these two results was 
within 0.01 mm\footnote{Cox claims \cite{Cox97} that in the ``step gauge experiment''
using ten zirconia reference spheres with sphericity $ 1 \, \mu m $ the maximum standart 
uncertainty of $ 0.75 \, \mu m $ was achived.}. Each 
result was obtained by 10 measurements on the Carl Zeiss Prismo 10 S-ACC with 
$ MPE_{E} $ = $ 1.7 + \frac{L}{350} \, \mu m $. 

The author will appreciate any 
third party help or ideas, especially ones from developers of the Calypso 
software, to move this experiment to a good theoretical ground. 
See \cite{GUM} for an in-depth assessment of the evaluation of 
coordinate-measuring machines uncertainties.

\section{Limitations}

Since two points on a plane define rotation and planar origin, the 
proposed method can be applied directly only to the planar repositioning 
of workpieces. To do three dimensional repositioning at least three 
registration points 
should be used. Reference spheres \cite{Cox97} or
cylinders will be preferable for this task instead of plates (fig.~\ref{Plate}). 
Spheres also can be probed from any direction and cylindrical holes only 
from a direction parallel to their axis. Repositioning with the usage of 
reference spheres in Calypso can be automated the same way.

Nevertheless, the described method of repositioning is significantly 
simpler and does not require any design changes to be done. Also it 
is possible to take advantage of available features on a part to create 
an intermediate datum.

\section{Conclusions}

It is easy to pass any number of parameters from
one alignment to another using the proposed method. The number of alignments 
is not constrained by two and can be as high as required 
to fulfill the measurement task completely. Coordinates, saved to the file, are
in Base alignment. They can be loaded from file at any time after the
original base alignment was changed due to repositioning of the workpiece.

The author was able to measure radii of large circular parts using this 
method. Dimensions of the parts were large enough to prevent them from 
arranging inside the CMM volume. It would be impossible to perform
this measurement on the available CMM without using the proposed method.

Application of this method is not limited to the flat large-scale parts only.
It can be applied to small ones if such parts have complex surfaces and 
require repositioning to be measured. No design changes to the parts are 
required.

The PCM approach is very easy to implement and it could be efficiently
used as common practice in Calypso measurements.

%\onecolumn

%\newpage

\section*{References}

%\section{Illustrations}
% \label{}

%\begin{center}
%   \includegraphics[width=8.5cm]{Eng3Fig.jpg} \\
%   {Fig.3. The complete measurement plan.}
%\end{center}
%\vspace{0.5cm}


% notes:

% \bibitem{label2}
% If there is a note, it should come last:
% \bibitem{label3} \note
% \end{subbibitems}

%\bibitem{}

%\bibliographystyle{IEEEtran}
\bibliography{ArticleKosarevsky}

\end{document}